\documentclass[pdflatex,sn-mathphys]{sn-jnl}

\usepackage[utf8]{inputenc}
\usepackage[numbers]{natbib}
\usepackage{multirow}
\usepackage{subcaption}
\usepackage{graphicx}
\usepackage{amsmath}
\usepackage{xcolor}
\usepackage{adjustbox}
\usepackage{comment}

\jyear{2023}%

\theoremstyle{thmstyleone}%
\theoremstyle{thmstyletwo}%

\theoremstyle{thmstylethree}%

\begin{document}

\title[A Comparison of Proximity Indices Applied to OpenAlex]{Measuring Technological Convergence in Encryption Technologies with Proximity Indices: A Text Mining and Bibliometric Analysis using OpenAlex}


\author*[1,2]{\fnm{Alessandro} \sur{Tavazzi}}\email{tavazale@gmail.com}

\author[1,3]{\fnm{Dimitri} \sur{Percia David}}

\author[1]{\fnm{Julian} \sur{Jang-Jaccard}}

\author[1]{\fnm{Alain} \sur{Mermoud}}

\affil[1]{\orgdiv{Cyber-Defence Campus}, \orgname{armasuisse Science and Technology}, \orgaddress{\street{Building I, EPFL Innovation Park}, \postcode{1015}, \city{Lausanne}, \country{Switzerland}}}

\affil[2]{\orgdiv{Institute of Mathematics}, \orgname{EPFL},\orgaddress{ \postcode{1015}, \city{Lausanne}, \country{Switzerland}}}

\affil[3]{\orgdiv{Institute of Entrepreneurship \& Management}, \orgname{University of Applied Sciences of Western Switzerland (HES-SO Valais-Wallis)}, \orgaddress{\street{Techno-Pôle 1, Le
Foyer}, \postcode{3960}, \city{Sierre}, \country{Switzerland}}}


\abstract{Identifying technological convergence among emerging technologies in cybersecurity is a crucial task for advancing science and fostering innovation. Unlike previous studies that focus on the binary relationship between a paper and the concept it attributes to technology, our approach utilizes attribution scores to enhance the relationships between research papers, combining keywords, citation rates, and collaboration status with specific technological concepts. The proposed method integrates text mining and bibliometric analyses to formulate and predict technological proximity indices for encryption technologies using the 'OpenAlex' catalog. Our case study findings highlight a significant convergence between blockchain and public-key cryptography, evident in the increasing proximity indices. These results offer valuable strategic insights for those contemplating investments in these domains.} 




\maketitle
\thispagestyle{empty}

\clearpage
\pagenumbering{arabic} 

\newpage 

\section{Introduction}\label{sec1}

In an era characterized by a technological revolution, understanding the dynamics of technological evolution, convergence, and emergence has become crucial for advancing science and fostering economic innovation \cite{zhou_unfolding_2019, wang_approach_2019, carley_indicator_2018, halaweh_emerging_2013}. While emergent technologies continue to reshape global landscapes socially, economically, and intellectually \cite{porter_emergence_2019}, a substantial gap exists in the literature concerning a comprehensive quantitative measure for assessing technological convergence \cite{curran_anticipating_2010, song_anticipation_2017}.

To address this critical gap, our study employs the integration of bibliometric indicators, such as collaboration, common keywords, and citations, to facilitate a multidimensional analysis of technological convergence. We leverage the OpenAlex database, a rich source of scholarly papers, to model the evolution of encryption technology from 2002 to 2022. Unlike previous studies that typically use the binary relationship between research and the concept attribution to a specific technology, we leverage OpenAlex's technology attribution scores. This enhances the relational granularity between research papers and specific technological concepts, thus improving the accuracy of identifying technological convergence.

Additionally, we use random forests to generate time series of proximity indices to forecast technological trajectories. This approach has allowed us to identify a significant convergence between blockchain and public-key cryptography, aligning well with the trend in the growing practical application of public-key cryptography within blockchain ecosystems.

By identifying early stages of technological convergence, our study not only complements but also extends traditional patent analyses, providing better insight into emerging technological trends. This foresight can guide strategic investment and development in cybersecurity to strengthen defenses against evolving cyber threats.

The paper is organized as follows. In Section \ref{literature_review}, we evaluate the strengths and weaknesses of existing literature. Section \ref{data} outlines the data processing methodology, and Section \ref{method} presents the details of our proposed model. The results and their interpretations are presented in Section \ref{results_and_discussion}. Finally, Section \ref{Conclusion} discusses the limitations of our study, potential extensions, and offers concluding remarks.

\section{Related Work}\label{literature_review}

This section provides an overview of the current state of the art in technology convergence approaches. Additionally, we explore various measurements and models that researchers have used to identify technological convergence, particularly in the context of emerging technologies.

\subsection{Technology Convergence Approaches}


The approaches used to assess technological convergence are diverse, with many researchers relying on proximity indices between technologies. While patent-based approaches are common in the literature (Song et al., 2017; Curran, 2011; Lee, 2018; Kim, 2017), our focus shifts towards a bibliometric-based approach. These approaches often utilize metrics like citation counts, publication frequencies, and keyword connections (Chand \& Bhatt, 2021). In addition, most existing research focuses on convergence after academic discovery but before market implementation (Curran, 2011). However, these studies often lack comprehensive early-stage scientific analysis, a gap we aim to address in our approach. 


Emerging technologies have a significant presence in the literature, with diverse definitions (Halaweh, 2013; Liu, 2020; Garner, 2017; Carley, 2018). We compare two main factors: one looks at socioeconomic factors like uncertainty and cost, while the other emphasizes scientific features such as novelty, growth, and community integration. In this study, we choose the latter, as it aligns with our focus on early-stage scientific convergence.

In existing works, technology convergence is typically studied separately from emerging technologies, often utilizing completely different sets of approaches. However, in our study, we aim to identify technology convergence in emerging technologies by combining and utilizing methods and indicators that are often employed separately


\subsection{Technology Convergence Measurements} \label{methods_models_lit}

In this section, we review the state of the art in bibliometric indicators, such as keywords, citations, and collaborations, along with our critiques. 

\subsubsection{Keywords}

Identifying common keywords across various technologies is considered a crucial measure of technological convergence. This approach, especially its variant focusing on dynamic networks formed based on shared keywords, has been central in several studies \cite{huang_detecting_nodate, song_discovering_2017, blouin_balkanized_2013, chand_bhatt_technology_2021, kim_machine-learning-based_2020}. Despite its usefulness, keyword analysis has limitations, including ambiguities from semantic similarities in common keywords and imprecision in individual keywords representing distinct concepts \cite{li_evolutionary_2016, chen_identifying_2016}. Nevertheless, our study assumes that a shared lexicon can be an important indicator of convergent objectives, implying technological convergence. This assumption aligns with the notion that a common vocabulary, such as keywords, typically signifies a shared understanding of interest in a certain technology \cite{cojocaru_role_2012}, thus widely accepted as an important measure \cite{huang_detecting_nodate, song_discovering_2017, blouin_balkanized_2013, chand_bhatt_technology_2021, kim_machine-learning-based_2020}.

\subsubsection{Citations}
Citation and co-citation analyses, serving as indicators of scholarly interdependence, are essential for understanding technological convergence \cite{li_evolutionary_2016, curran_anticipating_2010, he_synthetical_2022, klarin_systematic_2021}. However, these methods face challenges stemming from incomplete datasets and unreliable reference-listing algorithms \cite{curran_anticipating_2010}. Moreover, McRoberts' influential critique emphasized the limitations of citation analysis in fully capturing academic interconnectedness, pointing to unacknowledged influences, citation biases, literature ignorance, and self-citation practices as contributing factors \cite{macroberts_problems_1996}.

Despite these limitations, we make the assumption that citation metrics, while offering only a partial perspective, provide insights into knowledge exchange across various domains, indicating potential convergence \cite{curran_anticipating_2010, klarin_systematic_2021}. This perspective is supported by insights from Bröring, who suggests that interdisciplinary citations often signal the onset of scientific convergence, potentially leading to collaborative research ventures \cite{curran_anticipating_2010}.

\subsubsection{Collaborations}

In addition to keywords and citations, the research community recognizes other indicators of technological convergence. One such metric involves assessing the extent to which researchers contribute to multiple technological fields. We argue that this aspect of research collaborations plays a crucial role in driving scientific and technological advancements \cite{liu_structure_2015, curran_anticipating_2010}. From an organizational theory standpoint, collaborations signify the preliminary stages of technological convergence \cite{liu_structure_2015, curran_anticipating_2010}. As articulated by Bröring, an increase in collaborative efforts typically precedes the merging of distinct disciplines \cite{curran_anticipating_2010}. Therefore, analyses focusing on co-authorships and shared researcher pools hold value in predicting technological convergence.


\subsection{Technology Convergence Models}

This section explores graph-based and forecasting models used to identify and predict clusters of converging technologies from among other common technologies.

\subsubsection{Graph-based Models}
Graph-based strategies provide an alternative, allowing for an examination of technological convergence from a macro perspective, rather than relying solely on isolated pairwise analysis \cite{lu_link_2011, kim_anticipating_2019, breitzman_emerging_2015, huang_detecting_nodate, chand_bhatt_technology_2021, kumar_hypergraph_2020}. Cluster analyses and hypergraphs facilitate this process, aiding in the identification and prediction of converging technology clusters \cite{lu_link_2011, breitzman_emerging_2015, huang_detecting_nodate, chand_bhatt_technology_2021, kumar_hypergraph_2020}. This approach recognizes that technological convergence generally occurs within groups, highlighting the limitations of pairwise analyses in a field characterized by interconnected networks \cite{breitzman_emerging_2015, schoen_network_2012}. However, the broader perspective of graph-based methods may obscure subtle interactions at the micro-level, potentially oversimplifying or misrepresenting complex relationships and leading to information loss or biased interpretations.

\subsubsection{Forecasting Models}

Forecasting models for technological convergence leverage a diverse range of algorithms and methodologies. These models incorporate graph-based clustering algorithms such as Spectral Louvain Modularity (SLM) and the Louvain method, Design Structure Matrix (DSM) tools, random forests, and topological clustering to analyze the connections between technological evolutions and emergences \cite{huang_detecting_nodate, kim_forecasting_2017, kumar_hypergraph_2020, lee_early_2018, kim_machine-learning-based_2020, kajikawa_structure_2008}. In addition, predictive models utilize AutoRegressive Integrated Moving Average (ARIMA), neural networks, and exponential smoothing techniques to anticipate technological trends \cite{kim_forecasting_2017, mitrea_comparison_2009, lee_early_2018, kim_machine-learning-based_2020}.

\section{Data}\label{data}

This study utilizes data from OpenAlex, a comprehensive repository of research papers, authors, and institutions, which was created by drawing aspirations of the ancient Library of Alexandria \cite{priem_openalex_2022}. OpenAlex employs an automated system that assigns a set of Wikidata concepts to each paper from a pool of 65,000 concepts. This is accomplished through a multi-class deep learning classifier trained on the Microsoft Academic Graph. Each concept is assigned a score between 0 and 1, indicating the paper's relevance to that particular concept \cite{priem_openalex_2022}.

\vspace{5pt}
\textbf{Data Extraction and Processing}
\vspace{3pt}
\newline
Data relevant to the evolution of encryption technologies was extracted and processed into time series for analysis and forecasting. The focus was on scientific concepts related to 36 encryption technologies identified by experts through the Delphi method \cite{brown1968delphi}. Among the 36 technologies, only 25 were found in OpenAlex, forming the basis of our analysis. Papers not relevant to these 25 technologies were assigned a zero score.  
It was observed that over 90\% of the papers had a zero attribution score, affirming the importance of the scoring system (refer to Figure \ref{fig:dist_score_of_attribution} in the appendix).

\vspace{5pt}
\textbf{Data Refinement}
\vspace{3pt}
\newline
Due to the incomplete data in OpenAlex, papers without references, constituting half of the dataset, were removed to prevent bias. The distribution of these papers across technologies was uniform, ruling out any potential data skew (see Figure \ref{fig:dis_missing_ref} in the appendix). Additionally, we excluded another 5\% of the remaining papers that were not linked to any concepts. To address anomalies like the overrepresentation of papers published in January, we corrected them by evenly redistributing throughout the year. Duplicate entries were resolved by keeping the version with the most comprehensive information.

\vspace{5pt}
\textbf{Data Enhancement}
\vspace{3pt}
\newline
Keywords were assigned to each paper using KeyBert \footnote{https://maartengr.github.io/KeyBERT/api/keybert.html}, a model that identifies keywords based on their similarity to the text, utilizing 'cosine similarity' to indicate keyword relevance. Although a substantial portion of keywords was unique, common keywords played a crucial role in our analysis (see Figure \ref{fig:dis_keywords} in the appendix).

\vspace{5pt}
\textbf{Author Influence}
\vspace{3pt}
\newline
We calculated the h-index for authors in the encryption field from 2002 to 2022, generating both incremental and non-incremental indices on both a monthly and yearly basis. The incremental indices, offering a cumulative perspective of an author's impact, were considered a more accurate representation of an author's significance in the field (refer to Figures \ref{fig:dist_monthly_notincremental} and \ref{fig:dist_monthly_incremental} in the appendix).

\vspace{5pt}
\textbf{Data Consolidation}
\vspace{3pt}
\newline
The completed dataset included different elements like h-indices, references, and keywords. These were organized into a detailed data frame with columns such as 'paper,' 'keyword,' 'cosine similarity,' 'title,' 'publication date,' 'abstract,' 'year,' 'month,' 'author,' 'referenced works,' 'concepts,' 'score concepts,' 'yearly H index not incremental,' 'yearly H index incremental,' 'monthly H index incremental,' and 'monthly H index not incremental.' 


\section{Method}\label{method}
This section outlines the specifics of our proposed method, introducing novel mechanisms for calculating proximity indices using keywords, collaborations, and citations. Additionally, we delve into the details of how we fitted the data through interpolation to enhance the clustering and forecasting capabilities of the proposed method.

\subsection{Proximity Indices}\label{proximity_indices}
We aim to understand if certain indices capture specific stages of convergence more effectively than others. Additionally, we explore whether these indicators consistently align or may contradict each other in certain cases. Our hypothesis is that regardless of the indicator used to model technological convergence, they will generally reveal convergence between two technologies if it exists. However, this convergence may manifest differently or at different times. To investigate this, we create multiple indicators of technological convergence based on common keywords, citations, and collaboration between technologies.
 
We assign papers that are attributed to technologies with a score between 0 and 1. Similarly, computed keywords are assigned to papers with a 'cosine similarity' score between 0 and 1. To capture the importance of each author in the field of encryption technologies, we compute an artificial h-index specific to the set of papers under consideration. 

For each indicator, our approach involves incorporating all relevant information for each month of each year, assigning appropriate weights to the variables, and constructing a time series. For example, in the case of keywords, we compute all common keywords between two technologies for each month of each year. We then calculate the average number of occurrences of each keyword in papers related to these two technologies during that month. This value is multiplied by the average cosine similarity and the average score of attribution to the pairs of technologies of the papers where each keyword occurred. These weights reflect the importance of specific keywords and the pair of technologies at a given moment in time. Similar computations are performed for other indices.

It's important to note that the indices presented in this paper draw inspiration from the literature, where common keywords, collaboration, and citations are commonly used. However, the explicit form of the indices is novel, as they directly leverage a unique database attributing every paper to concepts with a score between 0 and 1.

Furthermore, we opt not to normalize the indices. Common normalization factors are either volatile, introducing bias, or yield very small indices, resulting in flat curves that hinder the identification of trends in technological convergence. The non-normalized time series of all indicators are presented on a single plot for all pairs of technologies.

Each index discussed below is computed for a specific month from 2002 to 2022. However, for simplicity, we denote the variables used in the computation of the indices without explicitly specifying their dependence on a month, even though these variables are inherently tied to a specific month.

\vspace{5pt}
\textbf{Keywords Proximity Index}
\vspace{3pt}

Let $t1$ and $t2$ be two technologies. Let $K_{t1-t2}$ denote a keyword: the keyword of some papers related to $t1$ and some papers related to $t2$ published during a specific month. Let $C^K_{t1-t2}$ be the average cosine similarity of all the times this keyword occurred as a keyword of a paper related to $t1$ or $t2$ for a specific month. This measures how significant the keyword is for the papers related to $t1$ or $t2$ for a specific month. Let $A^K_{t1-t2}$ be the average attribution score to $t1$ and $t2$ of all papers related to $t1$ or $t2$ published during a specific month, with $K$ as a keyword. This represents how close the papers having $K$ as a keyword are to the technologies $t1$ and $t2$ during a specific month. Let $N^K_{t1-t2}$ be the average of times $K$ occurs in papers related to $t1$ and to $t2$, which were published during a specific month.
For a given month $m$, we define the index of proximity based on keywords between two technologies $t1$ and $t2$ as follows:
$$ IK^m_{t1-t2} := \sum_{K_{t1-t2}}N^K_{t1-t2}*C^K_{t1-t2}*A^K_{t1-t2} $$
This index aims to measure how significant a keyword is for the proximity between a technology $t1$ and $t2$. The idea is, for each common keyword between both technologies, to multiply $C^K_{t1-t2}*A^K_{t1-t2}$, which works as a coefficient of importance for the keyword, with $N^K_{t1-t2}$ the number of times the keyword appears in papers related to both technologies. Then, we take the given sum over all common keywords to obtain a global degree of proximity. As a result, this index evaluates the proximity of technologies based on weighted common keywords present in both technologies.

\vspace{5pt}
\textbf{Citations Proximity Index}
\vspace{3pt}

Let $t1$ and $t2$ be two technologies. Let $C_{t1-t2}$ denote a paper related to the technology $t1$, which cites a paper related to the technology $t2$. Let $A^{C_{t1-t2}}_{t1}$ be the score of attribution to the technology $t1$ of the paper $C_{t1-t2}$ and $A^{C_{t1-t2}}_{t2}$ be the score of attribution to the technology $t2$ of the paper related to the technology $t2$ cited by the paper $C_{t1-t2}$. 

For a given month $m$, we define the two indices of proximity based on citations between two technologies, from $t1$ to $t2$ and from $t2$ to $t1$ as follows:
$$ IC^m_{t1-t2} := \sum_{C_{t1-t2}}(A^{C_{t1-t2}}_{t1}+A^{C_{t1-t2}}_{t2})*\frac{1}{2} $$
$$ IC^m_{t2-t1} := \sum_{C_{t2-t1}}(A^{C_{t2-t1}}_{t2}+A^{C_{t2-t1}}_{t1})*\frac{1}{2} $$
These indices measure the flow of citations from one technology to another during a given month. Furthermore, it assesses the importance of each citation for the connection between the two given technologies $t1$ and $t2$. As a result, these indices evaluate the proximity of technologies based on citations.

\vspace{5pt}
\textbf{Collaboration Proximity Index}
\vspace{3pt}

Let $t1$ and $t2$ be two technologies. Let $A_{t1-t2}$ denote an author who is the author of some papers related to $t1$ and some papers related to $t2$ published during a specific month. For an author $A_{t1-t2}$, let $H^{A_{t1-t2}}_n$ be its monthly not incremental h-index and $H^{A_{t1-t2}}_i$ be its monthly incremental h-index. 

The h-indices evaluate the influence of scientists by considering citations. In our case, h-indices enable a precise assessment of the significance of each author in establishing collaboration between two technologies. In fact, if an author has a high h-index, his impact on the collaboration between two technologies is assumed to be significant. 

Let $A^A_{t1-t2}$ be the average attribution score to $t1$ and to $t2$ of all the papers related to $t1$ or $t2$ published during a specific month having $A_{t1-t2}$ as an author. This quantifies the proximity of papers, for which $A_{t1-t2}$ serves as an author, to the technologies $t1$ and $t2$ within a designated month.
Let $N^A_{t1-t2}$ be the average of the times $A_{t1-t2}$ occurs in papers related to $t1$ and $t2$ published during a specific month. This variable significantly determines how relevant a given author's collaboration is to connecting two technologies. 

For a given month $m$, we define the two indices of proximity based on the collaboration of authors between two technologies, as follows.
$$ {IA^m_{t1-t2}}_n := \sum_{A_{t1-t2}}N^A_{t1-t2}*H^{A_{t1-t2}}_n*A^A_{t1-t2} $$
$$ {IA^m_{t1-t2}}_i := \sum_{A_{t1-t2}}N^A_{t1-t2}*H^{A_{t1-t2}}_i*A^A_{t1-t2} $$
These indices aim to measure how significant a collaboration is for the proximity between a technology $t1$ and $t2$. 
The approach we utilize involves the aggregation of coefficients assigned to each author, quantifying the significance and influence of their work within the connection between two technologies. This aggregation spans across all authors who have contributed to publications related to both technologies under consideration. As a result, this index evaluates the proximity of technologies based on collaboration.

\subsection{Interpolation and Fitting Data}

Once we've calculated all the proximity indices, the next step involves interpolating the time series of these indices. It was necessary because some technologies lack associated papers in certain months. Across the 625 time series, we achieve an average interpolation rate of around 20\%, a reasonable percentage given that some technologies have only a limited number of attributed papers.

For the interpolation process, we employ a polynomial method of degree 3, which is a mathematical technique widely used in data analysis for approximating a curve or function. 
In practical terms, polynomial interpolation utilizes a polynomial function of a specified degree to approximate a function passing through a set of discrete data points. For example, a cubic polynomial is employed for interpolation of degree 3. Given four distinct data points ($x_0$, $y_0$), ($x_1$, $y_1$), ($x_2$, $y_2$), and ($x_3$, $y_3$), the formula for estimating the value y at some intermediate value of x, denoted as $x_i$, using cubic polynomial interpolation is expressed as:
$$y = a_0 + a_1*(x_i - x_0) + a_2*(x_i - x_0)^2 + a_3*(x_i - x_0)^3$$
To determine the coefficients $a_0$, $a_1$, $a_2$, and $a_3$, a system of equations based on the given data points must be solved. Various methods, such as Lagrange interpolation or the Newton divided difference method, can be employed to derive these coefficients.

The simplest form of interpolation is linear interpolation, directly estimating values between two adjacent data points. This assumes a linear relationship between the data points. Specifically, for two points ($x_1$, $y_1$) and ($x_2$, $y_2$), the linear interpolation formula for estimating the value y at some intermediate value of x, denoted as $x_i$, is given by:
$$y = y_1 + (x_i - x_1) * (y_2 - y_1) / (x_2 - x_1)
$$
In practice, one can vary the degree of interpolation based on the data's nature and the desired accuracy. Linear interpolation (degree one) is quick and simple but assumes a linear relationship, while higher-degree polynomial interpolation provides a more exact fit through data points but may introduce unnecessary oscillations.
In our case, we initially chose a polynomial interpolation of degree 3 for the time series, offering flexibility without excessive oscillations. However, this leaves missing values at the time series' extremities, where some series start or end with gaps. To address this, we apply a second interpolation using a linear method.

This two-step process ensures all time series are filled throughout the entire period. Negative values created by the interpolation are replaced with zeroes, as defined by the indices, which can only yield non-negative values.

\begin{figure}[!htp]
    \centering
    \includegraphics[width=8cm]{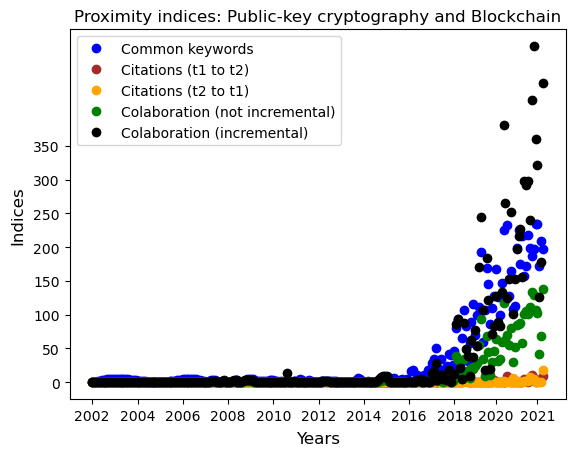}
    \caption{Indices of proximity between Public-key cryptography and Blockchain from 2002 to 2021.}
    \label{fig:scatterplot_indices}
\end{figure}

Upon plotting some of the time series, we observe a cloud of points moving in a certain direction rather than forming a proper line, as depicted in Figure \ref{fig:scatterplot_indices}.

The observed fluctuations in the time series can be attributed to various factors influencing our computations. These factors include computed variables such as cosine similarities or h-indices, as well as variables present in the OpenAlex dataset, like the attribution score to technologies based on the classification algorithm.
Moreover, the dynamic nature of the scientific landscape, subject to monthly variations, contributes significantly to the time series' fluctuations. This dynamism arises from various causes, including varying publication frequencies for different technologies and the sudden integration of new scientific discoveries or algorithms leading to substantial citations by a significant portion of the scientific community, among other influences. Numerous factors contribute to why the computed time series are not entirely smooth. Nevertheless, the crucial aspect of our work lies in the general tendencies reflected by these indices.

To address the variability, we opt to fit curves to the points obtained for each time series. We perform polynomial interpolation, computing eleven polynomials for each time series with degrees ranging from 0 to 10. To determine the best fit for the time series, we select the polynomial interpolation with the lowest Symmetric Mean Absolute Percentage Error (SMAPE). An illustrative example is provided in Figure \ref{fig:fitting_curves} below.
\begin{figure}[H]
\centering
    \includegraphics[width=8cm]{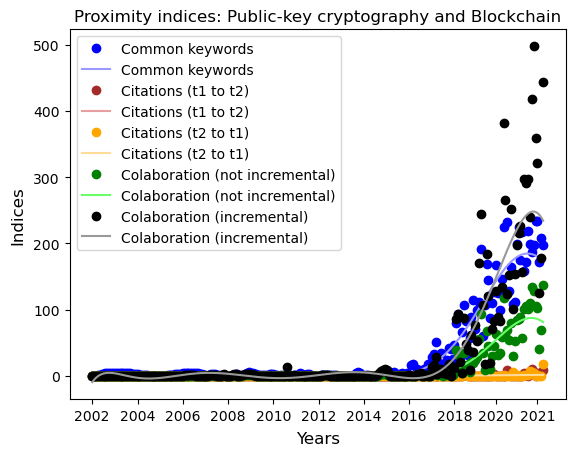}
    \caption{Optimal polynomial fitting of the time series of proximity indices between Public-key cryptography and Blockchain with an interpolation rate of 24\% from 2002 to 2021.}
    \label{fig:fitting_curves}
\end{figure}

\subsection{Clustering}

In this study, our focus is on implementing clustering methods to rearrange time series data that represent technological proximity based on their shapes across various index types. The objective of this analysis is to pinpoint time series exhibiting increasing proximity, indicating potential technological convergence. 

Our initial step involves preprocessing the time series. Following numerical experiments, we decide against deseasonalizing the time series, as this introduces significant negative values, which contradicts our non-negative indices' definition. Subsequently, we normalize all time series between 0 and 1. 
The normalization process begins with standardizing the time series data by subtracting the minimum value from each data point and then scaling the entire series by dividing it by the maximum value within the adjusted time series. This results in a time series that proportionally expresses the position between the minimum and maximum values of the original time series. Lastly, due to the high volatility of the normalized time series, we opt to smooth them using exponential smoothing with a robust smoothing parameter of $\alpha = 0.1$ using the following formula:


$$F_t = \alpha x_t + (1-\alpha) F_{t-1}\quad\text{, where}$$
\begin{itemize}
    \item $F_t$ is the forecast for the current time period $t$.
    \item $x_t$ is the actual observation in the current period.
    \item $F_{t-1}$ is the previous forecast.
    \item $\alpha$ is the smoothing parameter ($0 < \alpha < 1$).
\end{itemize}
The smoothing parameter $\alpha$ controls the weight given to the current observation versus the previous forecast. A smaller $\alpha$ assigns more weight to older data, while a larger $\alpha$
gives more weight to recent data.

Following the application of exponential smoothing, we exclude all-time series from our analyses with an interpolation rate exceeding 50\%. These time series, accounting for 25\% of the dataset, are considered artificial and represent noise in our dataset, potentially biasing our results.

Next, we allocate all normalized time series with a mean lower or equal to 0.02 to an artificial cluster representing a flat time series. This artificial cluster encompasses approximately 15\% of the entire time series set. This approach allows the clustering algorithm to focus exclusively on data that cannot be manually clustered, constituting around 60\% of our original time series.

We employ three clustering algorithms for time series: K-Means, K-Shape, and K-Medoids. 
We decided to fit these clustering algorithms in two ways: first, by training them on one part of our time series, and second, by training them on a large set of approximately 20'000 time series that we import from the Darts library \footnote{https://unit8co.github.io/darts/}. To measure the distance between time series, we use the metric ``l1'', the sum of the absolute value of the differences between vectors or time series, as in our case. We do not want to penalize time series with large numerical differences too much, as this would be the case with the Euclidean distance, where the difference between points is squared, since the main goal is to cluster them by shape and not directly by the values they take.

To evaluate the performance of the clustering algorithm, we implement several strategies. We first use the silhouette score \cite{rousseeuw1987silhouettes} to compute the clustering quality for each sample, each cluster, and finally for all the data. Then, we visualize the number of time series by cluster to see how the time series are distributed among the clusters. Afterward, we visualize the centroids of the clusters on a surface, with a positioning showing distances proportional to their distances as time series and with sizes proportional to the number of samples contained in each cluster. Finally, for each cluster, we plot randomly chosen time series within the cluster to assess their similarities visually.


To calculate silhouette score, one first calculates the average distance $a_i$ from $x_i$ for a given data point $x_i$ to all other data points within the same cluster. This measures the cohesion of the data point with its cluster.

Then, one calculates the average distance $b_i$ from $x_i$ to all data points in the nearest neighboring cluster (i.e., the cluster other than the one to which $x_i$ belongs). This measures the separation of the data point from other clusters.

The silhouette score $s_i$ for the data point $x_i$ is then calculated using the following formula:
$$s_i = (b_i - a_i) / \operatorname{max}(a_i, b_i)$$
If the value of $s_i$ is close to 1, it indicates that the data point is well-clustered and is far away from neighboring clusters. If the value of $s_i$ is close to -1, it suggests that the data point is misclassified into the wrong cluster, as its distance to its own cluster is much greater than the distance to the nearest neighboring cluster. If the value of $s_i$ is around 0, it means that the data point is on or very close to the boundary between two clusters. Regarding the overall silhouette score for an entire clustering solution (a set of clusters), one obtains it by computing the silhouette score for each data point in the dataset and then taking the average of these individual silhouette scores.


Ultimately, K-Shape with 5 clusters emerged as the optimal choice for our time series, achieving an average silhouette score of 0. This result indicates that the clusters are reasonably well-balanced, and the centroids exhibit sufficient separation from each other.
\vspace{5pt}


\subsection{Forecasting}\label{forecasting}
Our goal is to predict proximity indices and, ultimately, forecast how the closeness between encryption technologies evolves in order to predict technological convergence. We employ four methods to forecast time series data for 3, 6, and 12 months, aiming for optimal results.

In the first method, known as "local forecasting," each algorithm is trained on a segment of each time series, and then the given time series is directly forecasted with the fitted algorithm.

In the subsequent methods, "clustering forecasting" and "global forecasting," we categorize time series based on common keywords, citations, and collaborations. In clustering forecasting, each algorithm is trained on all-time series within a cluster, testing if using an algorithm trained on similar time series data leads to better forecasting. In global forecasting, each algorithm is trained on an all-time series, forecasting an all-time series with the fitted algorithm.

The last method, "transfer learning forecasting," involves training each algorithm on a large set of time series imported from Darts and then forecasting all-time series with the fitted algorithm.

Before forecasting, we preprocess the time series by cleaning the data to remove noise. We select time series with less than 50\% interpolation, smooth them using exponential smoothing with a smoothing parameter of $\alpha = 0.1$, and start forecasts from December 2021, considering incomplete updates for 2022 in OpenAlex data.



To understand forecasting quality, we use the Symmetric Mean Absolute Percentage Error (SMAPE).

Let $A:={\{A_t\}}_{t \geq 1}^n$ and $F:={\{F_t\}}_{t \geq 1}^n$ be respectively the actual and the forecasted time series taking values at $n$ time periods.
The symmetric mean absolute percentage error between $A$ and $F$ is defined as follows:
$$ \textbf{SMAPE}(A,F) := \frac{100}{n}\sum_{t=1} \frac{\lvert F_t - A_t \rvert}{(\lvert F_t \rvert +\lvert A_t \rvert)\frac{1}{2}} $$
If the actual and the forecasted time series are both zero at some time $t$, the contribution to the global sum is defined as zero. Using the triangle inequality, we obtain the following bound for SMAPE:
$$ \textbf{SMAPE}(A,F) \leq  200$$
since
$$ \textbf{SMAPE}(A,F) \leq \frac{100}{n}\sum_{t=1} \frac{\lvert F_t \rvert + \lvert A_t \rvert}{(\lvert F_t \rvert +\lvert A_t \rvert)\frac{1}{2}} = \frac{100}{n}\sum_{t=1} \frac{1}{\frac{1}{2}} = \frac{100}{n}\frac{n}{\frac{1}{2}} = 200 $$

It is noteworthy that if the actual time series consistently remains zero while the forecasted time series consistently stays non-zero, even if the predicted time series closely aligns with the actual one, SMAPE between both time series will be 200\%. Given the presence of numerous flat time series in our data, this observation should be kept in mind during our analysis of the results.

We employ k-fold cross-validation with an expanding window to assess forecasting quality. In this approach, each time series is divided into successive sections. Training windows are created, starting from the first section and expanding at each step with the next section. Algorithms are trained on each training window, with forecasts initiated from the end of each window. The SMAPE between the forecasts and actual values is computed for each iteration, and the average of these SMAPEs represents the global error for a specific algorithm. This method is employed to mitigate overfitting.

\begin{figure}[htp]
    \begin {flushleft}
    \includegraphics[width=12cm]{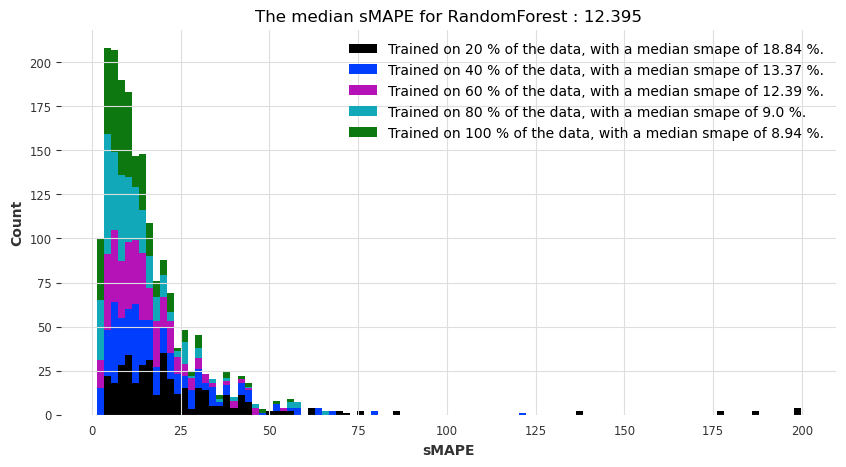}
    \caption{Distribution of forecasting errors obtained by predicting time series of keyword-based proximity indices. The forecasting algorithm employed is Random Forest, trained globally on all our time series, with a forecasting horizon of 6 months}
    \label{fig:crossvalid_ex}
    \end {flushleft}
\end{figure}

To visualize the distribution of forecasting errors across all expanding windows, we generate a histogram illustrating the number of forecasts by error size in a single plot with distinct colors, as depicted in Figure \ref{fig:crossvalid_ex}.

More concretely, we split the data into several sets. We take 80\% of the data to train our algorithm, and 20\% is left for testing the algorithm. Then, out of our training set, we take 80\% of the data to tune the hyperparameters of our models and 20\% to validate the results. We optimize the hyperparameters of each algorithm for each type of forecasting. Then, we train the models and forecast the time series using the k-fold cross-validation with an expanding window, as explained above. Last, to compare the different performances between the algorithms for each forecasting horizon and type of forecasting, we visualize the trade-off between the error and the computation time. This allows us to choose the optimal method for each specific forecasting task, as discussed in Section \ref{discussion_results}.

\section{Results and Discussion}\label{results_and_discussion}

This section delves into the details of the results and their insights, followed by a case study illustrating the effectiveness of our proposed method in two cybersecurity topics. Finally, we discuss the limitations and future work.

\subsection{Overall Results}\label{discussion_results}

The following tables provide detailed forecasting accuracy for each method, presenting data specific to index types, algorithms, and forecasting horizons. In these tables, Index 1 and Index 2 represent citations from one technology to another and vice versa. Index 3 and 4 denote collaboration based on incremental and non-incremental h-indices, respectively, while Index 5 pertains to common keywords. The colors \textcolor{brown}{brown}, \textcolor{violet}{violet}, and \textcolor{blue}{blue} indicate the best forecasting accuracy for horizons of \textbf{3}, \textbf{6}, and \textbf{12} months, respectively.

\begin{table}[!htb]
\begin{center}
\begin{adjustbox}{width=9.9cm,center}
\begin{minipage}{1\textwidth}
\caption{Median SMAPEs for local forecasting}\label{local_forecasting}%
\begin{tabular}{@{}lllllll@{}}
\toprule
Algorithm & Horizon & Index 1 &	Index 2	& Index 3 &	Index 4 &	Index 5\\
\midrule
\multirow{3}{*}{Naive seasonal (K=1)}  & 3 months & 
3.89 & 3.89 & 20.95 &	20.82 &	\textcolor{brown}{6.70}   \\
 & 6 months & 5.37& 5.37 & 	35.09 &	33.54 &	11.29  \\
 & 12 months & 8.28 & 8.28 &	\textcolor{blue}{54.56} &	50.78 &	14.97 \\
\hline
\multirow{3}{*}{Naive seasonal (K=12)}  & 3 months & 3.02 & 3.02 &	88.89 &	80.78 &	28.38   \\
& 6 months & 5.24 & 5.24 & 87.75 &	82.94 &	28.17  \\
 & 12 months & 9.66	 & 9.66	 & 86.11	 & 81.75	 & 26.63\\
\hline
\multirow{3}{*}{Exponential smoothing}  & 3 months  &	\textcolor{brown}{1.53}&	\textcolor{brown}{1.53}&	20.39& 20.22&	7.05 \\
& 6 months & \textcolor{violet}{3.59} & \textcolor{violet}{3.59} & 	33.56 &	31.64 &	11.36  \\
 & 12 months & 10.10 & 10.10 &	57.66 &	54.18 &	15.81  \\
\hline
\multirow{3}{*}{Theta}  & 3 months & 1.99 & 1.99 &	\textcolor{brown}{19.91} &	\textcolor{brown}{20.06} &	7.08  \\
& 6 months & 4.47 & 4.47 &	\textcolor{violet}{32.98} &	\textcolor{violet}{30.96} &	\textcolor{violet}{11.07} \\
 & 12 months & \textcolor{blue}{8.18}  & \textcolor{blue}{8.18} &	54.97 &	\textcolor{blue}{50.69} &	\textcolor{blue}{14.50} \\
\hline
\botrule
\end{tabular}
\end{minipage}
\end{adjustbox}
\end{center}
\end{table}

\begin{table}[!htb]
\begin{center}
\begin{adjustbox}{width=9.9cm,center}
\begin{minipage}{1\textwidth}
\caption{Median SMAPEs for randomized clustering forecasting}\label{rand_cluster_forecasting}%
\begin{tabular}{@{}lllllll@{}}
\toprule
Algorithm & Horizon & Index 1 &	Index 2	& Index 3 &	Index 4 &	Index 5 \\
\midrule
\multirow{3}{*}{Linear Regression}  & 3 months &
74.09 &	33.65 &	116.38 &	128.49 &	28.36  \\
 & 6 months & 37.84 &	45.83 &	76.63 &	62.15 &	21.22 \\
 & 12 months & 32.31 &	26.54 &	78.06 &	69.87 &	17.95 \\
\hline
\multirow{3}{*}{LGBM}  & 3 months & 36.73 &	32.32 & 73.20 &	79.81 &	11.99   \\
& 6 months & 38.91 &	30.48 &	48.72 &	51.47 &	11.61  \\
 & 12 months & 24.94 &	20.94 &	53.04 &	44.72 &	13.44  \\
\hline
\multirow{3}{*}{Random Forest}  & 3 months & \textcolor{brown}{3.06} &	\textcolor{brown}{2.36} &	\textcolor{brown}{19.41} &	\textcolor{brown}{19.01} &	\textcolor{brown}{6.55} \\
& 6 months & \textcolor{violet}{4.19} &	\textcolor{violet}{2.46} &	\textcolor{violet}{28.80} &	\textcolor{violet}{27.10} & \textcolor{violet}{9.55} \\
 & 12 months & \textcolor{blue}{2.97} & \textcolor{blue}{5.55} &	\textcolor{blue}{43.02} &	\textcolor{blue}{37.12} &	\textcolor{blue}{11.97}  \\
\hline
\botrule
\end{tabular}
\end{minipage}
\end{adjustbox}
\end{center}
\end{table}

\begin{table}[!htb]
\begin{center}
\begin{adjustbox}{width=9.9cm,center}
\begin{minipage}{1\textwidth}
\caption{Median SMAPEs for non-randomized clustering forecasting}\label{cluster_forecasting}%
\begin{tabular}{@{}lllllll@{}}
\toprule
Algorithm &  Horizon & Index 1 &	Index 2	& Index 3 &	Index 4 &	Index 5 \\
\midrule
\multirow{3}{*}{Linear Regression}  & 3 months &
41.31 &	39.45 &	84.32 &	98.18 &	17.97\\
 & 6 months &  75.34 &	35.25 &	82.19 &	82.75 &	16.42\\
 & 12 months & 59.14 &	45.08 &	100.35 &	89.80 &	19.96 \\
\hline
\multirow{3}{*}{LGBM}  & 3 months & 59.17 &	15.89 &	80.05 &	108.49 &	17.44\\
& 6 months & 49.16 & 15.47 &	70.92 &	84.61 & 	18.46\\
 & 12 months & 50.32 &	21.86 &	56.16 &	85.09 &	14.36 \\
\hline
\multirow{3}{*}{Random Forest}  & 3 months &  \textcolor{brown}{8.24} &	\textcolor{brown}{4.47} &	\textcolor{brown}{41.61} &	\textcolor{brown}{31.41} &	\textcolor{brown}{8.26} \\
& 6 months & \textcolor{violet}{8.26} &	\textcolor{violet}{6.05} &	\textcolor{violet}{43.24} &	\textcolor{violet}{33.06} &	\textcolor{violet}{10.47} \\
 & 12 months &  \textcolor{blue}{11.94} &	\textcolor{blue}{5.11} &	\textcolor{blue}{65.71} &	\textcolor{blue}{38.29} &	\textcolor{blue}{12.08}  \\
\hline
\botrule
\end{tabular}
\end{minipage}
\end{adjustbox}
\end{center}
\end{table}

\begin{table}[!htb]
\begin{center}
\begin{adjustbox}{width=9.9cm,center}
\begin{minipage}{1\textwidth}
\caption{Median SMAPEs for global forecasting}\label{global_forecasting}%
\begin{tabular}{@{}lllllll@{}}
\toprule
Algorithm & Horizon & Index 1 &	Index 2	& Index 3 &	Index 4 &	Index 5 \\
\midrule
\multirow{3}{*}{Linear Regression}  & 3 months &
\textcolor{brown}{0.0} & 	\textcolor{brown}{0.0}	 &  62.04 & 	51.85 & 	10.87  \\
 & 6 months &  \textcolor{violet}{0.0} & 	\textcolor{violet}{0.0}	&  70.17 & 	63.72 & 14.08\\
 & 12 months & \textcolor{blue}{0.0} & 	\textcolor{blue}{0.0} & 	86.31 & 	90.55 & 	17.10 \\
\hline
\multirow{3}{*}{LGBM}  & 3 months &0.0 & 	0.0 & 	31.37 & 	33.17	&  10.22 \\
& 6 months & 0.0& 	0.0& 	47.61 & 	45.59	& 13.19 \\
 & 12 months & 0.0 & 	0.0& 	59.60 & 	52.98 & 	16.09 \\
\hline
\multirow{3}{*}{Random Forest}  & 3 months & 0.0& 	0.0 & 	\textcolor{brown}{22.01} & 	\textcolor{brown}{21.62} & 	\textcolor{brown}{8.25} \\
& 6 months & 0.0 & 	0.0	&  \textcolor{violet}{39.27} & 	\textcolor{violet}{35.07} & 	\textcolor{violet}{11.63}  \\
 & 12 months & 0.0 & 0.0 & 	\textcolor{blue}{54.55} & 	\textcolor{blue}{49.82} & 	\textcolor{blue}{16.08}  \\
\hline
\botrule
\end{tabular}
\end{minipage}
\end{adjustbox}
\end{center}
\end{table}

\begin{table}[!htb]
\begin{center}
\begin{adjustbox}{width=9.9cm,center}
\begin{minipage}{1\textwidth}
\caption{Median SMAPEs for transfer learning forecasting}\label{transferlearning_forecasting}%
\begin{tabular}{@{}lllllll@{}}
\toprule
Algorithm & Horizon & Index 1 &	Index 2	& Index 3 &	Index 4 &	Index 5 \\
\midrule
\multirow{3}{*}{Linear Regression}  & 3 months &
\textcolor{brown}{91.06} & 	\textcolor{brown}{90.19}	 &  \textcolor{brown}{107.99} & 	\textcolor{brown}{128.38} & 	\textcolor{brown}{82.52}  \\
 & 6 months &  \textcolor{violet}{92.06} & 	\textcolor{violet}{91.44}	&  \textcolor{violet}{118.68} & 	\textcolor{violet}{135.36} & \textcolor{violet}{102.58}\\
 & 12 months & \textcolor{blue}{94.80} & 	\textcolor{blue}{94.32} & 	\textcolor{blue}{128.14} & 	\textcolor{blue}{144.67} & 	\textcolor{blue}{117.81}\\
\hline
\multirow{3}{*}{LGBM}  & 3 months & 98.47 & 	97.93 & 	165.79 & 	171.97	&  182.81 \\
& 6 months & 98.50 & 	97.93 & 	166.26 & 	172.05	& 183.46 \\
 & 12 months & 98.62 & 	98.08 & 	167.35 & 	172.94 & 	184.71 \\
\hline
\multirow{3}{*}{Random Forest}  & 3 months & 98.18 & 	97.62 & 	164.16 & 	170.19 & 	182.06 \\
& 6 months & 98.03 & 	97.70	&  161.82 & 	170.404 & 	178.98  \\
 & 12 months & 98.31 & 97.67 & 	160.16 & 	169.59 & 	180.76  \\
\hline
\botrule
\end{tabular}
\end{minipage}
\end{adjustbox}
\end{center}
\end{table}

\vspace{20pt}
\newpage

Overall, no single forecasting method consistently outperformed others across all tasks. For local forecasting, exponential smoothing and Theta algorithms demonstrated superior performance, while random forest outperformed linear regression and light gradient boosting machine (LGBM) in other forecasting types. In contrast, transfer learning methods exhibited comparatively poorer performance, potentially due to their computationally intensive nature and associated training limitations.

Different index types exhibited diverse forecasting outcomes. Citation indices, often flat, were straightforward to predict, resulting in a median SMAPEs of 0 across all forecasting horizons (see Table \ref{global_forecasting}). Conversely, collaboration indices posed greater challenges, with the most accurate predictions stemming from randomized clustering forecasting using random forests (refer to Table \ref{rand_cluster_forecasting}). Keyword indices saw reasonable forecasting, with algorithmic clustering via random forests providing the most accurate results (Table \ref{cluster_forecasting}).

\subsection{Case Study}\label{case_study}

\begin{figure}[htp]
\centering
    \includegraphics[width=10cm]{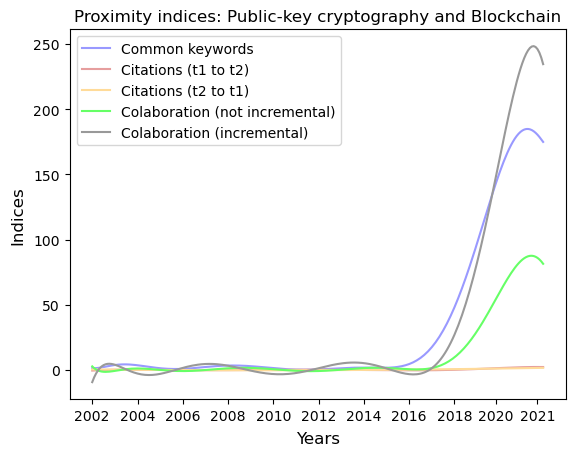}
    \caption{Case study: Examining the technological proximity between public-key cryptography and blockchain. The proximity indices undergo interpolation and are modeled using polynomial curves.}
    \label{fig:use_case}
\end{figure}

To illustrate the practical implications of our findings, we examine the evolution of proximity indices between "Public-key cryptography" and "Blockchain," as depicted in Figure \ref{fig:use_case}.

A notable interpolation rate of 46\% is evident. This can be attributed to the sparse interactions between both technologies from 2002 to 2012, reflecting the early stages of blockchain development in the cybersecurity domain. Consequently, during this phase, indices remained relatively stagnant. It's important to note that this interpolation rate, though steep, doesn't significantly bias our indices between "Public-key cryptography" and "Blockchain"; instead, it accurately reflects the non-interactive phase between these technologies.

A significant surge in the proximity indices is evident from 2017 onwards, with the indices for common keywords and collaboration based on incremental h-indices being particularly prominent. This suggests a robust correlation between the variables "common keywords" and "collaboration," indicating a notable number of authors simultaneously exploring both public-key cryptography and blockchain technologies. Additionally, there is a marginal increase in mutual citations between the two technologies.
While citation-based indices exhibit modest growth, it could be attributed to their inherent construction rather than a lack of interaction between the two technologies. Based on this evidence, we hypothesize that a tangible technological convergence occurred between blockchain and public-key cryptography from 2017 to 2021.

This convergence is highlighted by the increasing adoption of public-key cryptography in blockchain platforms during this period. Techniques like digital signatures, based on public-key cryptography, became integral to verifying blockchain transactions. Furthermore, the integration of zero-knowledge proofs, also based on public-key cryptography, gained traction in blockchain frameworks to validate assertions without revealing the underlying data. In conclusion, our findings highlight a pivotal phase of technological convergence between blockchain and public-key cryptography spanning 2017 to 2021.

\subsection{Limitations and Future Works}

While our research offers valuable insights, it also has limitations and suggests areas for future work. One limitation of our proposed method is its reliance on a single data source, OpenAlex, without integrating or comparing it to other similar datasets. This could potentially result in our proposed method being overfitted towards OpenAlex. Additionally, our approach lacks a normalization strategy for various calculations of proximity indices. As a result, index comparisons may be influenced by predominant values, such as the high number of publications in a specific technology domain. To address this issue, we propose index normalization by averaging monthly publications relevant to the technologies or using a weighted factor based on available variables.

Introducing a normalized index as an additional graph edge could enhance the computation of different proximity indices. This additional edge, forming 5-dimensional edges instead of our quintet of distinct indices, could contribute to establishing a more improved convergence threshold. Subsequently, leveraging community detection algorithms might provide insights into technological convergence by identifying clusters of converging technologies. This avenue holds promise for further exploration.

\section{Conclusion}\label{Conclusion}

In this study, we introduce a method that uses text mining and bibliometrics techniques, utilizing the "OpenAlex" catalog, to create and predict technological proximity indices specific to encryption technologies. A case study applying our method highlights the convergence between blockchain and public-key cryptography. The insights gained from our method can offer valuable guidance for stakeholders and enthusiasts navigating the impacts of encryption technology.

We recognize certain constraints in our study. Our literature corpus was limited to "OpenAlex," which may not comprehensively represent the core interests and activities in the research community working on encryption technologies. Additionally, the non-normalized nature of our indices hinders direct comparisons among various data sources. These limitations also suggest potential paths for future research to expand our findings, enriching the field of scientific monitoring within the encryption technology community.

\vspace{5pt}
\backmatter

\bmhead{Supplementary information}

The code of this project is on \url{https://github.com/technometrics-lab/Proximity-indices-applied-to-OpenAlex}. 

\bmhead{Acknowledgments}
We thank Jacques Roitel, whose work was the original base of this paper. Additionally, we would like to thank the other interns working at armasuisse at the time this research was led for their helpful coding advice: Léo Meynent, Marc Egli, Francesco Intoci, Lucas Crijns, and Alexander Glavackij.

\bibliography{myreferences}

\newpage

\begin{appendices}

\section*{Appendix} \label{appendixA}

This appendix includes figures, graphics, and bar charts that were referenced but not directly included in the main paper. They are presented in the order of their mention in the paper.

\begin{figure}[H]
\centering
    \includegraphics[width=8cm]{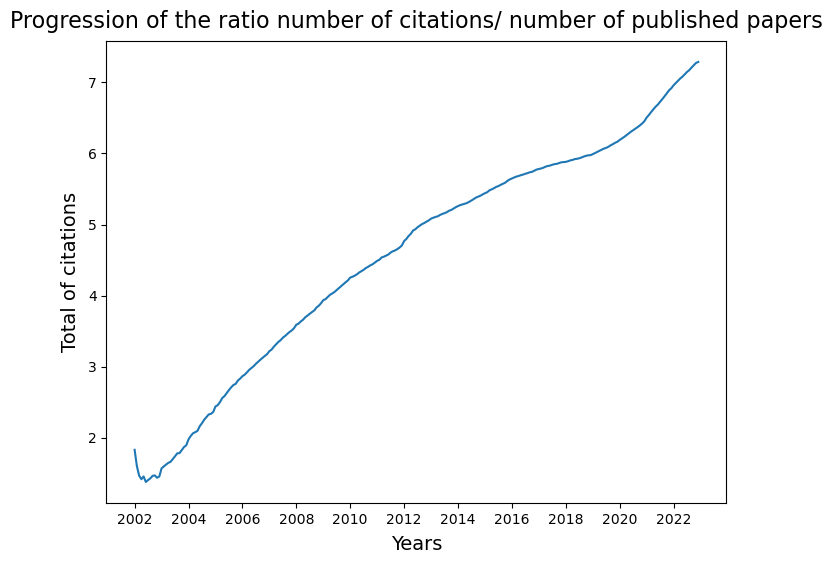}
    \caption{Evolution of the ratio between the number of citations and the number of published papers in the field of encryption technologies.}
    \label{fig:ratio_cit}
    \end{figure}

\begin{figure}[H]
    \begin {flushleft}
    \includegraphics[width=12cm]{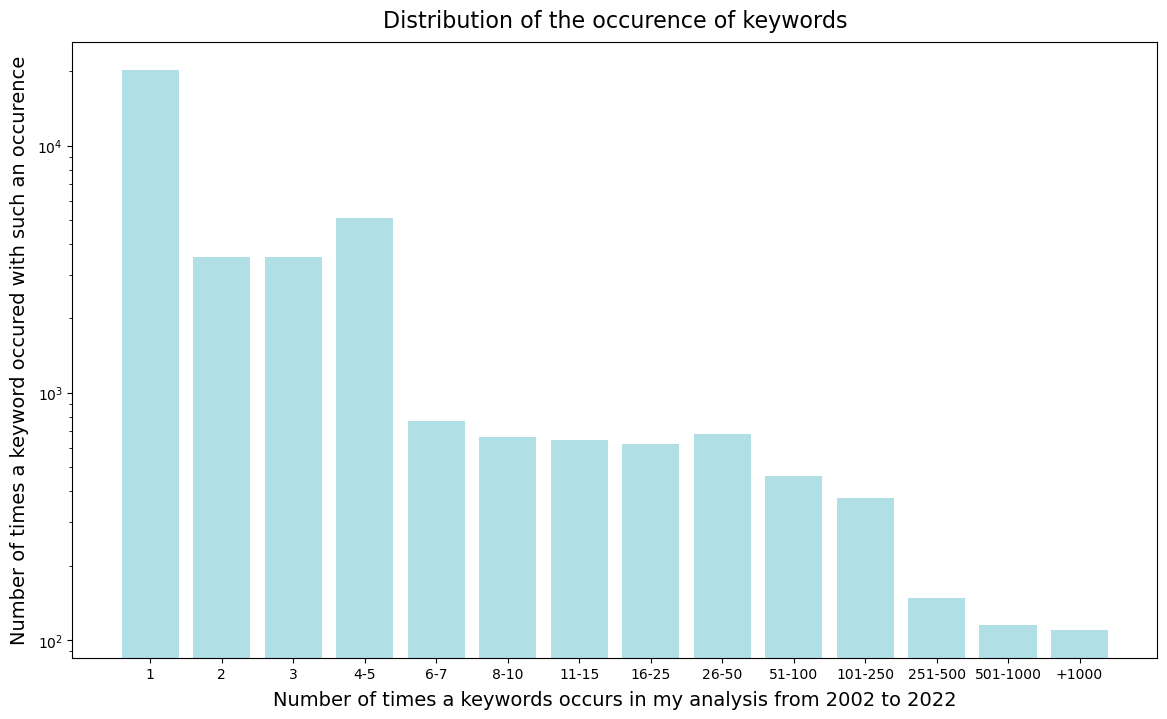}
    \caption{Distribution of the occurrence of keywords among all papers related
    to encryption technologies published between 2002 and 2022.}
    \label{fig:dis_keywords}
    \end {flushleft}
\end{figure}

\begin{figure}[H]
    \centering
    \includegraphics[width=10.5cm]{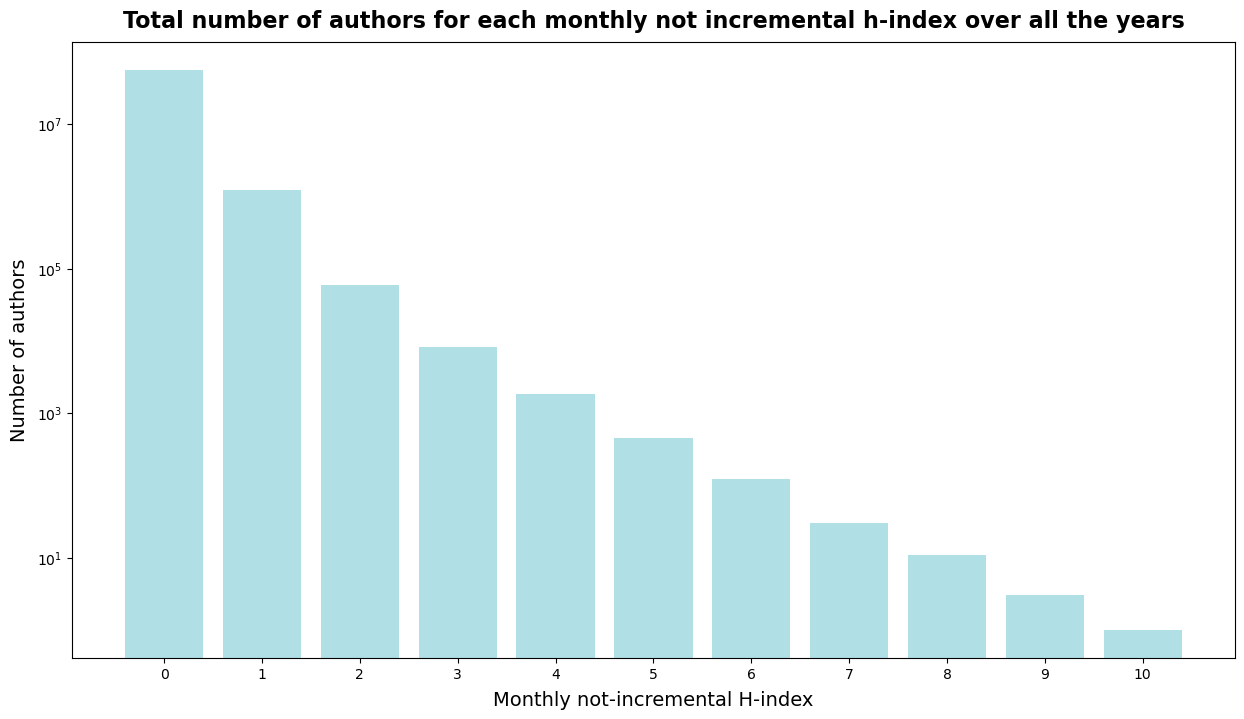}
    \caption{Distribution of the monthly non-incremental h-indices of all the authors of papers related to encryption technologies between 2002 and 2022.}
    \label{fig:dist_monthly_notincremental}
\end{figure}

\begin{figure}[H]
\centering
    \includegraphics[width=12cm]{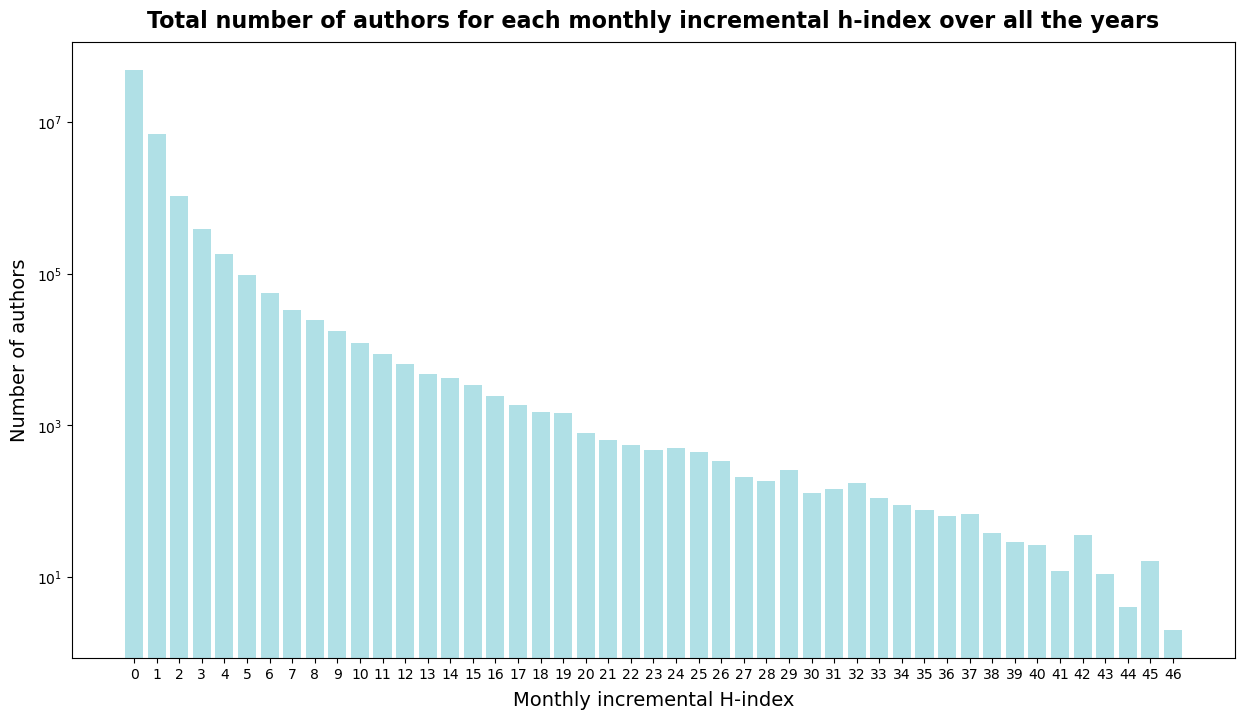}
    \caption{Distribution of the monthly incremental h-indices of all the authors of papers related to encryption technologies between 2002 and 2022.}
    \label{fig:dist_monthly_incremental}
    \end{figure}
    
\begin{figure}[H]
  \centering
    \includegraphics[width=10cm]{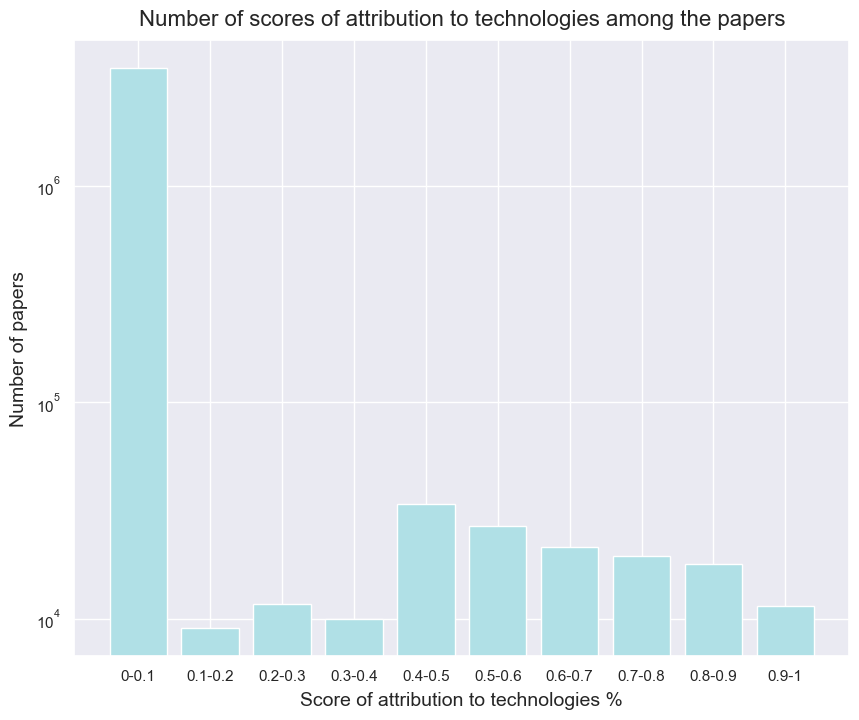}
    \caption{Distribution of the score of attribution to concepts given by OpenAlex among all papers related to encryption technologies published between 2002 and 2022.}
    \label{fig:dist_score_of_attribution}
\end{figure}

\begin{figure}[H]
\begin {flushleft} \includegraphics [width=11.5cm]{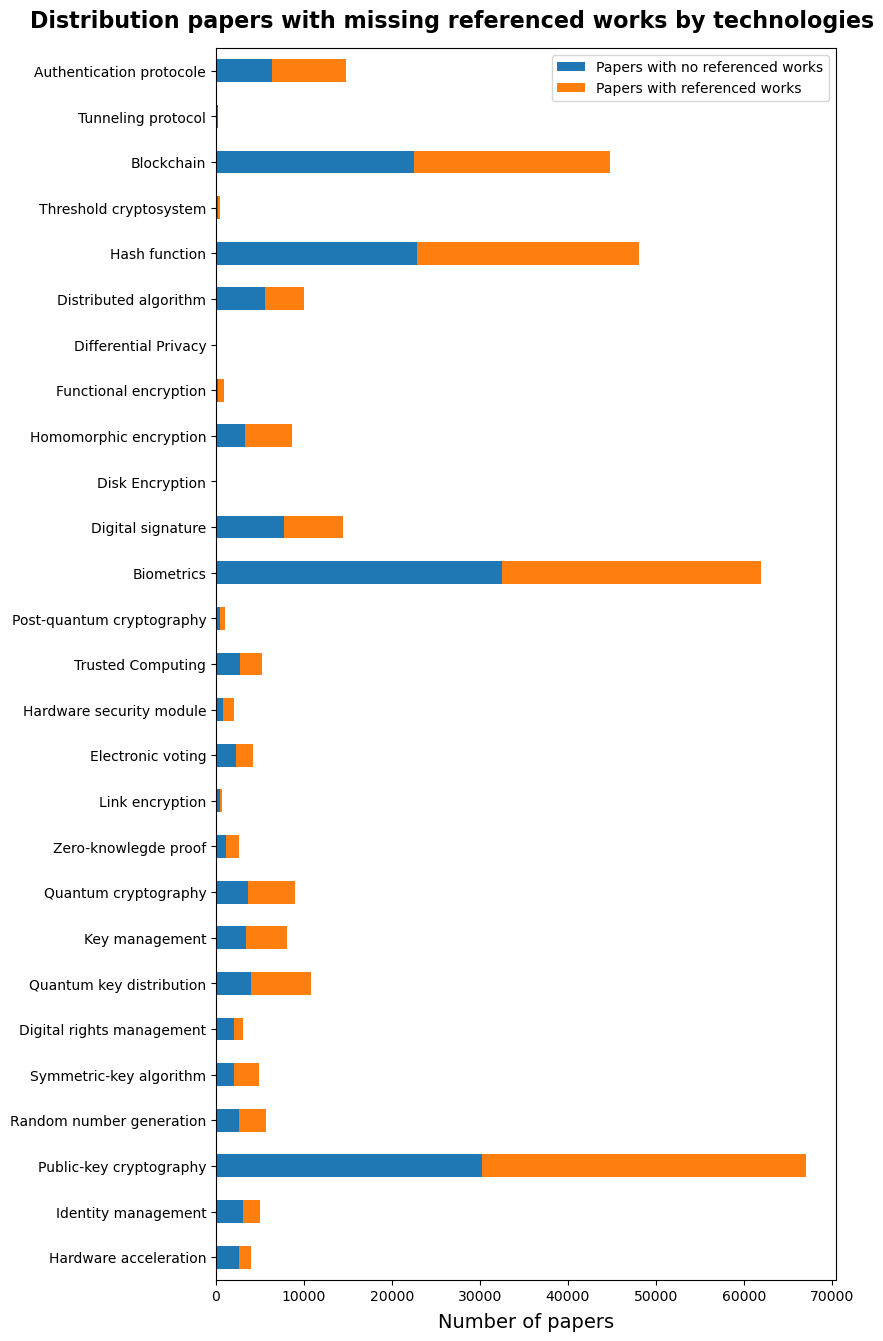}
    \caption{Distribution of the papers with missing referenced works by technologies for all the papers related to encryption technologies published between 2002 and 2022.}
    \label{fig:dis_missing_ref}
    \end {flushleft} 

\end{figure}


\end{appendices}


\end{document}